\documentclass[aps,prl,reprint,superscriptaddress,floatfix]{revtex4-2}
\usepackage{graphicx}
\usepackage{amsmath}
\usepackage{amssymb}
\usepackage{mathrsfs}
\usepackage[T1]{fontenc}
\usepackage{calligra}
\usepackage{array}
\usepackage{xcolor}
\usepackage{float}
\usepackage{soul}
\usepackage{bm}
\usepackage[toc,page]{appendix}
\usepackage{braket}
\usepackage{wrapfig}
\DeclareUnicodeCharacter{2212}{-}

\usepackage[colorlinks=true, linkcolor=blue, urlcolor=blue,
            citecolor=blue]{hyperref}

\hypersetup{
 colorlinks,linkcolor=blue,citecolor=blue,urlcolor=blue}

\begin{document}

    \title{Designer spin-orbit superlattices: symmetry-protected Dirac cones
    and spin Berry curvature in two-dimensional van der Waals metamaterials}
    \author{L. M. Martelo}
    \email{martelo@fe.up.pt}
    \affiliation{Departamento de Engenharia F\'isica, Faculdade de Engenharia da Universidade
    do Porto, Rua Dr. Roberto Frias, 4200-465 Porto, Portugal}
    \affiliation{Centro de F\'isica do Porto, Rua do Campo Alegre, 4169-007 Porto, Portugal}
    \author{Aires Ferreira}
    \email{aires.ferreira@york.ac.uk}
    \affiliation{School of Physics, Engineering and Technology and York Centre for Quantum Technologies, University of York, YO10 5DD, York, United Kingdom}


    \begin{abstract}

    The emergence of strong relativistic spin-orbit effects in low-dimensional systems provides a rich opportunity for exploring unconventional states of matter. Here, we present a route to realise tunable relativistic band structures based on the lateral patterning of proximity-induced spin-orbit coupling. The concept is illustrated on a patterned graphene--transition metal dichalcogenide heterostructure, where the spatially periodic spin-orbit coupling induces a rich mini-band structure featuring massless and massive Dirac bands carrying large spin Berry curvature. The envisaged systems support robust and gate-tunable spin Hall responses driven by the quantum geometry of mini-bands, which can be tailored through metasurface fabrication methods and twisting effects. These findings open pathways to two-dimensional quantum material design and low-power spintronic applications.

    \end{abstract}
    \maketitle


    The intertwining of orbital and spin degrees of freedom underpin a wealth of phenomena, from the formation of topological insulators to the spin Hall effect of light \cite{SOC_Liberman_1992,SOC_Onoda_2004,SOC_Kane_2005,SOC_Bernevig_2006,SOC_Berard_2006,SOC_Bliokh_2008,SOC_Bliokh_2010}. In condensed matter systems, spin-orbit coupling (SOC) is a relativistic interaction due to the motion of electrons in the electric field of the crystal lattice, which can yield spin-dependent band structures and Berry-curvature effects that strongly influence the electrodynamics of quasiparticles \cite{xiao_berry_2010,Resta_11}. Because the Berry curvature flux encodes global topological invariants (such as the Chern number for quantum anomalous Hall insulators), SOC is also a key mechanism behind quantised transport in topological phases of matter \cite{RevModPhys.83.1057_Xiao-Liang,RevModPhys.95.011002_CuiZu}. 

    Broken symmetries alter the spin-orbital character of electronic states \cite{Dresselhaus_55,Samokhin_09}, and therefore provide pathways by which to realise novel spin phenomena. Among these, the emergence of spin textures in spin-orbit-coupled systems with broken spatial inversion symmetry has generated enormous excitement in the fields of spintronics and magnonics recently \cite{Soumyanarayanan_2016,Yu_21}. Owing to a close interplay of spin, lattice (pseudospin), and orbital degrees of freedom, SOC manifests both in real and momentum spaces---spin-momentum locking of spin-split Fermi surfaces \cite{Ishizaka_11,Offidani_17,Feng_19}, magnetic skyrmions \cite{Bogdanov_94,bode_chiral_2007,muhlbauer_skyrmion_2009}, and persistent spin helices \cite{Bernevig_06,Koralek_09,RevModPhys.89.011001_Schliemann} are prominent examples---and forms the basis of several transport effects of fundamental and practical interest. Chief among these is the current-driven spin polarisation that occurs in non-magnetic conductors with nontrivial spin textures, such as spin-momentum-locked Rashba interfaces and topological surfaces \cite{Sanchez_16,Lesne_16,Kondou_16}. The ensuing net spin polarisations are often large (allowing current-induced magnetisation switching of ferromagnets  \cite{Mellnik_14,Yabin_14,Wang_15}) and tend to lie perpendicularly to the applied electric field owing to the tangential nature of conventional Rashba-type spin textures. Moreover,  recent studies have found that the net spin orientation can be tuned in chiral materials  boasting more exotic spin textures due to fully broken reflection symmetries \cite{Acost_21,Furukawa_21,Veneri_22,Seungjun_22}, which has the potential to unlock unconventional spin-orbit torques \cite{MacNeill_17,Sousa_20,Gosh_23}.  

    Likewise, the rich landscape of spin Hall effects (SHEs)  reflects the symmetries underlying spin-orbit-coupled matter \cite{Sinova_15}. Of recent and growing interest is the SHE in vertical heterostructures built from graphene and two-dimensional (2D) semiconductors  \cite{Sierra_21}. In these systems, the interfacial breaking of point-group symmetries leads to two main types of SOC that can be either induced or greatly enhanced via proximity effects: the sublattice-staggered SOC (underlying the valley-Zeeman effect) and the more familiar Rashba SOC \cite{Pachoud_14,Kochan_17,Perkins_24}. Beyond featuring an exceptionally high degree of SOC tunability via  strain and twisting effects \cite{Alessandro_19,Li_19,Rao_23,Sun_23,Tiwari2022,Fulop_21}, proximitized 2D crystals support robust extrinsic SHEs due to scalar impurities, having no counterpart in other, non-Dirac 2D systems \cite{milletari_covariant_2017,Perkins_SHE_2024} (for a recent review see Ref. \cite{Perkins_24}). Such symmetry-breaking effects are also of ubiquitous importance for 2D quantised transport \cite{Tang_17,Wu_18,Garcia_20,Zhao_21}, as well as for metallic anomalous Hall and magnetic spin Hall phases  \cite{Zhenhua_10,Offidani_18}.   
 
\begin{figure}
    \begin{centering}
    \includegraphics[width=1\columnwidth]{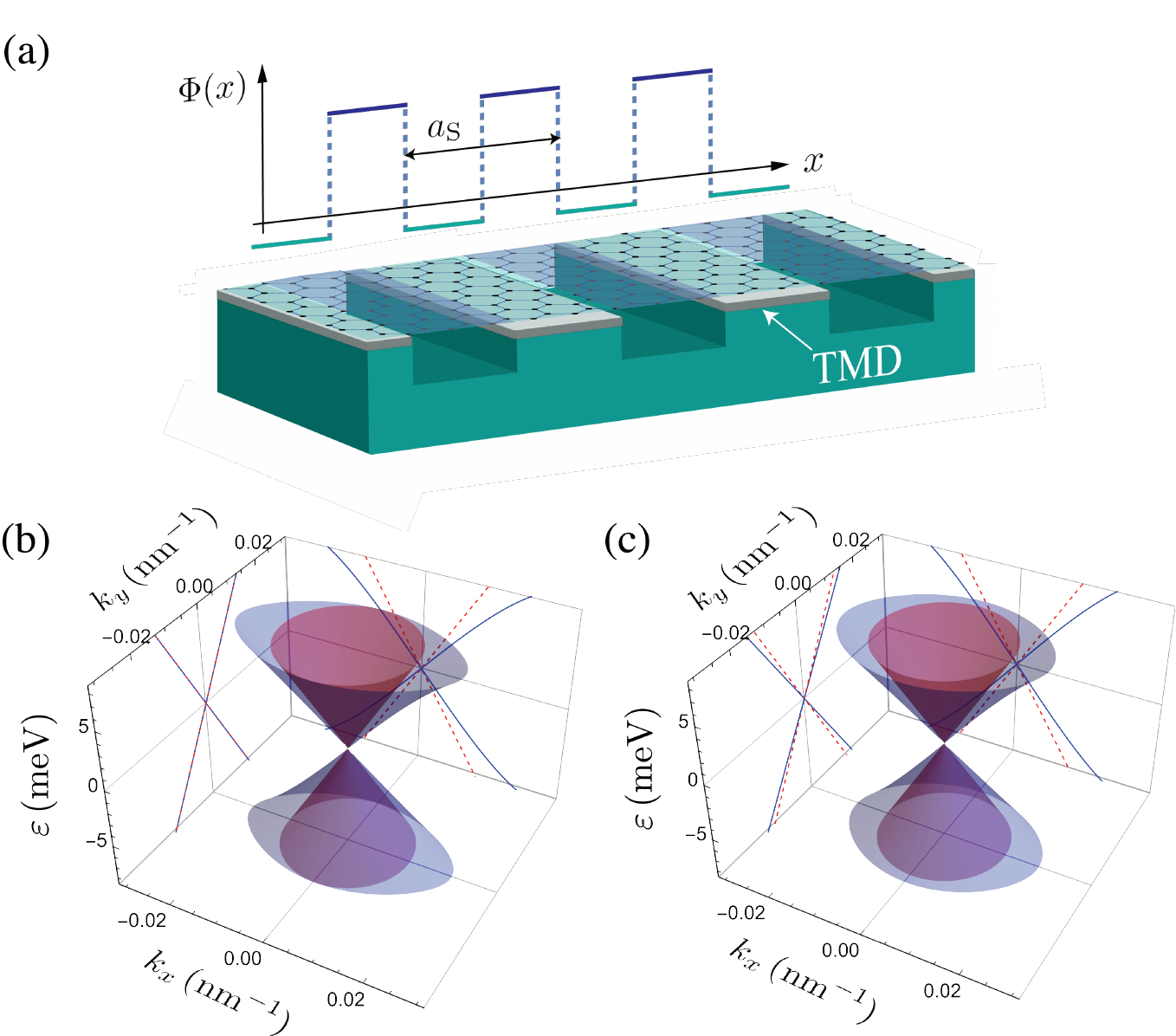}
    \par\end{centering}
    \caption{\textbf{\label{fig:01}}(a) 1D periodic modulation of the proximity-induced SOC. In this example, the SSOF is imprinted on graphene via the use of a dielectric metasurface decorated with 2D semiconductors (labelled TMD). (b) Energy dispersion of low-lying states around the $K$  valley for a zero-mean square-wave profile with $a_{\text{S}}=100$~nm, $\lambda_{\text{KM}}=20$~meV, and $u=0$. (c) Same as (b) but with $u=15$ meV. As a guide to the eye, the bare energy dispersion of graphene is shown in red (inner cones).}
    \end{figure}

    Despite this, most theoretical work so far has focused on translation invariant spin-orbit fields that reflect the periodicity of the underlying crystal structure, since this is the most conspicuous case. An interesting exception is the modulation of the strength of Rashba and Dresselhaus SOC induced in quantum-wire setups, previously explored in the context of spin-transistor devices \cite{Waveguide_1,Waveguide_2,Waveguide_3}. Inspired by recent advances in the realisation of artificial Dirac band structures in graphene with 
 one-dimensional (1D) superlattice potentials \cite{SL_Park_08,SL_Forsythe_18,SL_Li_21}, the purpose of this work is to show that the quantum geometry and electrodynamic response of 2D materials can be engineered via synthetic spin-orbit fields created by a metasurface.  Our proposal, outlined in Fig. \ref{fig:01}(a), leverages proximity-induced effects between atomically thin crystals to engender effective spin-orbit fields with periodicity $a_{\text{S}}$ much greater than the lattice scale, which we call super-spin-orbit fields (SSOFs). We envision that the long-wavelength modulation of the spin-orbit field acting on charge carriers can be achieved by placing graphene on a patterned high-SOC substrate, akin to the patterning of electrostatic potentials in a lateral graphene superlattice \cite{SL_Park_08,SL_Forsythe_18,SL_Li_21} (other possibilities are discussed below). As we shall see, the envisaged synthetic SSOFs not only lead to the formation of mini-bands but also impact their underlying quantum geometry, yielding a number of useful effects. This includes the counterintuitive and exotic possibility of creating linearly dispersing spin-degenerate electronic states even for Rashba-type SSOFs where spatial inversion symmetry is strictly broken. Our proposal is, therefore, complementary to previous superlattice setups, where the presence of spatially uniform SOC components generally leads to spin-split energy bands with non-linear dispersion as well as energy gaps  \cite{Lenz_2011,SuperlatticeUniformSOC_Li17,SuperlatticeUniformSOC_Mughnetsyan19}. The SSOFs are also distinct from superlattices arising from the periodic modulation of on-site staggered potentials \cite{Zarenia_12,Martino_23}, which lack SOC effects. Another advantage of the SSOFs introduced here is that they intrinsically generate semimetallic phases without the need for periodic Zeeman fields \cite{SuperlatticeUniformSOC_Brey16}. Moreover, similar to spatially uniform SOC, the SSOFs endow the electronic states of graphene with spin Berry curvature, thus paving the way to SHEs with unique geometric features.
 

    \textit{Setting the scene}.---To model the electronic properties of a graphene sheet subject to a proximity-induced SSOF, we employ a continuum low-energy description based on the Weyl-Dirac Hamiltonian \cite{peres_colloquium_2010},  supplemented with a 1D periodic perturbation comprising a scalar potential $U(x)$ \cite{SL_Park_08} and  SOC terms allowed by symmetry \cite{Pachoud_14,Kochan_17,Perkins_24}.  We focus exclusively on long-period perturbations,
hence suppressing intervalley scattering \cite{SL_Park_08}.  The Hamiltonian in the valley-isotropic basis is 
    \begin{equation}
          H_\tau=v\,(\boldsymbol{\sigma}\cdot\mathbf{p})\otimes s_0 +
        U(x)\sigma_{0}\otimes s_{0}+H_{\text{so},\tau}(x),
            \label{eq:Hamiltonian}
    \end{equation}
    where $v$ is the bare Fermi velocity of 2D massless Dirac fermions ($v=10^{6}$~m/s), $\sigma_a$ and  $s_a$ ($a=x,y,z$) are Pauli matrices acting on the pseudospin and spin subspaces, respectively, $\sigma_0$ and $s_0$ are $2\times 2$ identity matrices, $\mathbf{p}=-i\hbar\boldsymbol{\nabla}$ is the momentum operator, and $\tau = \pm 1$  is the valley index.  For the broad class of Dirac Hamiltonians that are locally invariant under the $C_{3v}$ point group \cite{Kochan_17,Perkins_24}, the SSOF term $H_{\text{so},\tau}(x)$ receives up to 3 contributions, namely, a spin-flip Rashba term [$H_{R}(x)=\lambda_{\textrm{R}}\,\Phi(x)\,(\sigma_{x}\otimes s_{y}-\sigma_{y}\otimes s_{x})$], a valley-Zeeman term due the broken sublattice symmetry [$H_{\textrm{vz},\tau}(x)=\tau\lambda_{\textrm{vz}}\,\Phi(x)\,\sigma_{0}\otimes s_{z}$], and a Kane-Mele (KM) term [$H_{\textrm{KM}}(x)=\lambda_{\textrm{KM}}\,\Phi(x)\,\sigma_{z}\otimes s_{z}$]. Here, $\lambda_{\textrm{R}}$, $\lambda_{\textrm{vz}}$ and $\lambda_{\textrm{KM}}$ respectively denote the nominal strength of the Rashba, valley-Zeeman and intrinsic-like SOC  induced by the application of the SSOF, while $\Phi(x)$ describes the spatial profile of the SOC modulation [$\Phi(x+m\,a_{\text{S}})=\Phi(x)$ with $m$ an integer]. We note that the presence of the periodic modulation yields a mini-band energy spectrum, ${\varepsilon_{n\mathbf{k}}}$, where $n\in\mathbb{Z}\backslash\{0\}$ is the mini-band index and  $\mathbf{k}$ is the Bloch wavevector relative to a Dirac valley; see Supplementary Material (SM) appended below for additional details.

    An example of a graphene system subject to a SSOF with a square-wave profile is depicted in Fig. \ref{fig:01}(a).  Recent measurements \cite{Tiwari2022,Rao_23,Sun_23} have shown that proximity-induced Rashba SOC in graphene/WSe$_{2}$ attains giant values of up to $15$ meV \cite{Sun_23},  which is more than 350 times larger than graphene's intrinsic SOC \cite{Sichau_19}, and makes group VI dichalcogenides ideal high-SOC substrates for our proposal. Lastly, we assume that the superlattice potential, when present, is designed to track the SSOF modulation (e.g. via a patterned bottom gate), and thus write $U(x)=u \, \Phi(x)$, where $u$ is the scalar potential amplitude. The most striking scenario, on which we will focus our attention, concerns Rashba SSOFs with a zero-mean profile [that is, $\langle\Phi(x)\rangle=(1/a_{\text{S}})\int_{0}^{a_{\text{S}}}dx\,\Phi(x)=0$]. For example, this can be accomplished through the encapsulation of a graphene sheet between identical dielectric layers with a relative offset of $a_{\text{S}}/2$. More exotic experimental routes, yet viable, include  metal intercalation   \cite{calleja_spatial_2015}, periodic folding of graphene \cite{costa_origami-based_2013}, proximity coupling to rippled group-VI dichalcogenides \cite{Castellanos-Gomez_13,Luo_2015} and deposition of graphene on stepped surfaces \cite{ortega_electron_2000}. The low-energy physics in all these routes are captured by Eq. (\ref{eq:Hamiltonian}) (or simple generalisations thereof) with a suitable choice of parameters. Without loss of generality, we work within the {$K$ valley} {($\tau=1$)} with a valley-degeneracy factor of two properly accounted for in physical quantities like the spin Hall conductivity.
    
    
    \textit{Results}.---To build intuition, we first consider an SSOF with zero spatial average ($\langle \Phi(x) \rangle =0$) that locally preserves all the spatial symmetries of the honeycomb lattice, i.e. with a single term ($H_{\text{KM}}$). The energy spectrum is two-fold spin degenerate in this case and exhibits the typical mini-band structure due to a synthetic periodic perturbation. In Fig. \ref{fig:01}(b), we show numerically exact results for a long-period square-wave modulation of type KM; details are provided in the  Methods. 
The most striking feature of the low-energy spectrum is the band touching at zero energy, i.e. the SOC spatial modulation precludes the opening of a topological gap \cite{SOC_Kane_2005}. (Higher-energy mini-bands are located at energies $\approx \pm 2 \pi v \hbar / a_{\text{S}} \approx \pm 40$ meV and thus lie outside the energy range of Fig. \ref{fig:01}.) Importantly, the linearly dispersing zero energy states in our system cannot be gapped out without breaking the global average condition of the periodic perturbation. In other words, the Dirac point degeneracy survives SSOFs with  $\langle \Phi(x) \rangle =0$ (the reader is referred to SM for numerical and analytical evidence supporting the generality of this statement). What is more, the emergent 2D Dirac fermions unveiled here remain massless even for SSOFs that locally break one or more spatial symmetries, such as a spatially modulated Rashba SOC. The robustness of the crossing point between the electron and hole mini-bands hints at significant effects of quantum geometry, which will be discussed shortly. 
    
    Next, we observe that the  KM-SSOF renormalises the group velocity along the modulation direction, $\hat x$, while it produces no change perpendicularly to it. This is the opposite behaviour of graphene under a periodic (scalar) potential  \cite{SL_Park_08}, and provides a simple mechanism to fine tune charge carrier propagation. This possibility is highlighted in Fig.~\ref{fig:01}(c), showing that the combined action of a SSOF and a periodic potential  squeezes the Dirac cones along both parallel and perpendicular directions to the reciprocal superlattice vector. Perturbation theory provides further insights, as detailed in the SM. In the limit $u,\lambda_{\text{KM}}\ll\hbar v/a_{\textrm{S}}$, the component of the group velocity parallel to the wavevector $\textbf{k}$ of the low-lying Dirac states reads as
        
    \begin{equation}
    v_{\hat{\mathbf{k}}} \cong v\left[1-\xi\,\frac{u^{2}\sin^{2}\theta_{\mathbf{k}}+\lambda_{\text{KM}}^{2}\cos^{2}\theta_{\mathbf{k}}}{\hbar^{2}v^{2} {G_1}^{2}}\right], 
    \label{eq:group-velocity}
    \end{equation}
    
  where $\theta_\mathbf{k}$ is the wavevector angle, $\xi$ is a geometric factor ($\xi\approx1.645$ for a square-wave SSOF), and $G_1 = 2\pi /a_{\text{S}}$. Equation (\ref{eq:group-velocity}) shows that the periodic perturbation can be tuned to yield an  isotropic group velocity. Indeed, setting $u=\pm\lambda_{\text{KM}}$ results in isotropic Dirac cones, thus mimicking the low-energy physics of bare graphene  without SOC. The situation becomes richer when considering realistic systems with broken spatial symmetries as shown below. For example,  Dirac fermions with isotropic behaviour can be engineered by means of a pure Rashba SSOF, bypassing the need for a scalar periodic potential. 
  
    \begin{figure}
    \begin{centering}
    \includegraphics[width=0.8\columnwidth]{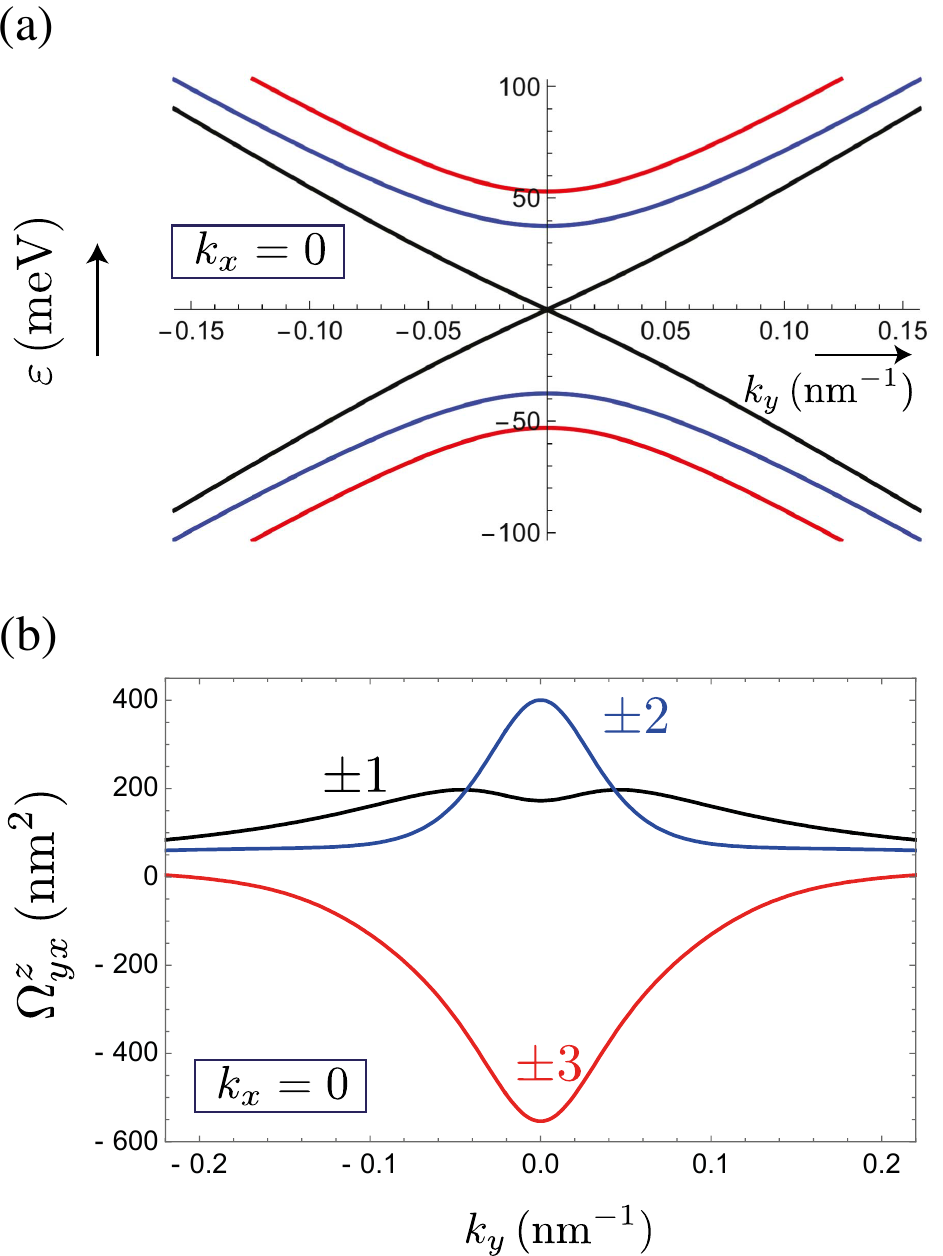}
    \par
    \end{centering}
    \centering{}
    \caption{\label{fig:02} (a) Dispersion of low-lying mini-bands along a cut with $k_{x}=0$.
    (b) Spin Berry  curvature along the same $\textbf{k}$-path.
    Mini-bands are labelled by integers next to curves {[}positive
    (negative) $n$ labels conduction (valence) bands{]}. 
    Other parameters: $a_{\text{S}}=100$ nm and $\lambda_{\text{R}}=20$ meV.}
    \end{figure}

    \begin{figure*}
    \centering{}\includegraphics[width=1\linewidth]{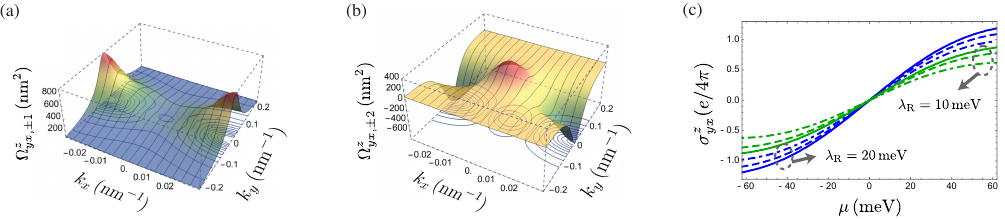}
    \caption{\label{fig:03} 
     (a) Momentum-space distribution of the SB curvature of the massless Dirac mini-bands ($n=\pm 1$). (b) Same as in (a) for the mini-bands $n=\pm 2$. SSOF parameters as in Fig. \ref{fig:02}. (c) Spin Hall conductivity $\sigma^z_{yx}$ as a function of the chemical potential for $k_B T = 26$ meV and selected twisting-induced SOC modulations, namely: pure Rashba SSOF (solid lines), admixed Rashba-valley-Zeeman SSOFs with $\lambda_{\text{vz}}=\lambda_{\text{R}}/2$ (dashed lines) and $\lambda_{\text{vz}}=\lambda_{\text{R}}$  (dot-dashed lines). Here, green (blue) curves correspond to $\lambda_R=10$ meV ($20$ meV).}
    \end{figure*}

    \textit{Realistic SSOFs and quantum geometry effects}.---Now, we turn to the class of SSOFs that admix valley-Zeeman ($H_{\textrm{vz}}$) and Rashba ($H_{\textrm{R}}$) terms due to the breaking of spatial symmetries. Unlike the KM-type SOC  in the example above, both $\lambda_{\textrm{R}}$ and $\lambda_{\textrm{vz}}$ can reach experimentally relevant energy scales, which is ideal for our proposal. We primarily focus on pure Rashba SSOFs which can be realised via twist-angle engineering in graphene-on-transition metal dichalcogenide (TMD) heterostructures \cite{Alessandro_19,Li_19,Rao_23,Sun_23}. We neglect the {KM}-type {SOC}, which due to its smallness \cite{Sichau_19} is unimportant.
 The idea is to tune the twist angle, so that the effective SOC of charge carriers on $A$ and $B$ sublattices coincide, yielding a vanishing valley-Zeeman effect, $\lambda_{\text{vz}}=(\lambda_{A}-\lambda_{B})/2=0$. The resulting SOC is thus of Rashba type (allowed by the broken $z\rightarrow -z$ symmetry) \cite{Alessandro_19,Li_19}.  This intriguing possibility has been confirmed experimentally via quasiparticle interference imaging \cite{Sun_23}, showing that $\lambda_{\text{vz}}\approx 0$ and $\lambda_{R} \approx 15$ meV for $30^\circ$ twist-angle graphene-on-WSe$_2$ systems. Armed with this important insight, we start by investigating the electronic structure induced by a square-wave Rashba SSOF. The energy dispersion of charge carriers in the three lowest-lying bands, above and below the charge neutrality point, is shown in Fig. \ref{fig:02}(a). The calculated spectrum contains several genuine fingerprints of the SSOFs proposed in this work. Similar to the case above, the zero energy modes exhibit linear dispersion (i.e. the Dirac point degeneracy is protected). Furthermore, the behaviour is isotropic. Thus, with regards to energy dispersion, this system emulates pristine graphene with a strongly renormalised Fermi velocity (see below). The massless nature of low-energy excitations is a robust feature of the 2D van der Waals metamaterials underpinning the SSOFs. In fact, only perturbations breaking the zero-average condition ($\langle \Phi \rangle=0$) can gap out the massless Dirac states (see SM). As such, the zero-energy modes can be mode as robust as desired in a realistic setup, by ensuring that the fabrication method preserves the global average of the periodic perturbation. This confers protection against local SOC fluctuations that are unavoidable in realistic systems.
      
Next, we ask whether the SSOFs can endow 2D massless Dirac fermions with quantum geometric properties. We start by noting that the mini-bands due to a square-wave Rashba SSOF [see Fig. \ref{fig:02}(a)] are two-fold spin degenerate, thus lacking a spin texture of their own. This is intriguing because the Rashba SOC breaks the spatial inversion symmetry and thus can lead to spin splittings. To explain this counter-intuitive result, we analytically compute the dispersion of the low-lying Dirac states using perturbation theory. While a standard second-order expansion in $\lambda_{\text{R}}$ predicts a spin-degenerate spectrum, a cumbersome third-order calculation yields     

\begin{equation}
\varepsilon_{n=\pm 1,{\bf k}s}^{(3)}\approx \pm \left(\hbar v_{\textrm{ren}}|{\bf k}|\,+s\tilde{\xi}\frac{\lambda_{\text{R}}^{3}}{\hbar^{2}v^{2}G_1^{4}}|{\bf k}|^{2}\right),   
\end{equation} 
where $s=\pm 1$ for spin-up (spin-down) low-energy branch, $v_{\textrm{ren}}=v[1-\xi(\lambda_{\textrm{R}}/\hbar vG_{1})^{2}]$  and $\tilde \xi$ is a geometric factor that equals zero for sine- and square-wave modulations, but is otherwise non-zero (e.g, for Kronig-Penney-type modulations,  $|\tilde \xi|$ attains values close to $0.26$; see SM for additional details. Hence, Rashba SSOFs with more general profiles can lift the spin degeneracy (as intuition would suggest), but only perturbatively. While the resulting spin splittings are typically small, a sizeable effect can be achieved by combining SSOFs with a periodic potential, providing a rich phenomenology for future exploration.

To examine the quantum geometry of SSOF-induced mini-bands, we map out the momentum-space distribution of the spin Berry (SB) curvature 
    
\begin{equation}
\Omega_{n}^{z}(\mathbf{k})=-2\hbar^{2}\,\text{Im}\;\sum_{m\neq n}\frac{\langle m\mathbf{k}|v_{x}|n\mathbf{k}\rangle\langle n\mathbf{k}|v_{y}^{z}|m\mathbf{k}\rangle}{(\varepsilon_{n\mathbf{k}}-\varepsilon_{m\mathbf{k}})^{2}},
\label{eq:spinBerry}
\end{equation}
    
where $v_i = v \, \sigma_i \otimes s_0$ and $v_i^z = v \, \sigma_i \otimes s_z$ (here, $i=x,y$) are the charge and spin velocity operators, respectively. This quantity governs the spin Hall transport of electron wavepackets and therefore is the geometric analogue of the Berry curvature in the anomalous Hall effect \cite{RevModPhys.82.1539_Nagaosa,Offidani_18}. The  SB curvature around the Dirac points is shown in Fig. \ref{fig:02}(b). We see that the linearly dispersing zero energy modes (labelled $n=\pm1$) are endowed with  significant SB curvature, despite their massless nature. This is evidently at variance with 2D gapless Dirac systems, which have vanishing (charge) Berry curvature \cite{RevModPhys.83.1057_Xiao-Liang}. Thus, the emergent 2D Dirac cones reported here are not only robust against perturbations  sharing the global average of the SSOFs but also display quantum geometric effects. To explore this further, Figs. \ref{fig:03}(a)-(b) show 3D plots of the SB curvature of mini-bands with $n=\pm 1,\pm 2$ in the mini-Brillouin zone. Two features are of note. First, the central peaks in the SB curvature of the massive Dirac mini-bands $n=\pm 2,\pm 3$ discussed earlier are seen to arise from local hot spots of SB curvature [see the local maximum of $\Omega_{yx,\pm2}^z$ at $\textbf{k}=0$ in Fig. \ref{fig:03}(b)]. The SB curvature also displays pronounced local minima near the zone edges, where $\Omega_{yx,\pm 2}$ attains large negative values. Second, the massless mini-bands ($n=\pm 1$) have a giant SB curvature at the edges of the mini-Brillouin zone ($k_x=\pm \pi/a_{\text{S}}$) [see Fig. \ref{fig:03}(a)], about twice as large as the Dirac-point hot spot of the massive mini-bands. We attribute this curious feature to the emergence of large pseudo-gaps along the SSOF direction; see SM. Finally, we note that the general behaviour is highly anisotropic, except in the immediate vicinity of the Dirac point.
     
 The enhanced SB curvature of the SSOF-induced mini-bands indicates that the 2D van der Waals metamaterials proposed here support large spin Hall responses. To confirm this, we compute the intrinsic spin Hall conductivity ($\sigma_{ij}^z$) from the flux of SB curvature using standard methods \cite{xiao_berry_2010} (see also Methods).
According to linear response theory, the $z$-polarised   spin current density generated by an external electric field is $\mathbf{j}_s^{z}=\sum_{i,j=x,y}\sigma_{ij}^{z}E_{j}\mathbf{e}_{j}$, where $E_j$ are the field components and $\textbf{e}_i$ is the unit vector along the $i$-axis. As shown in Fig. \ref{fig:03}(c), the spin Hall response has a strong energy dependence and can reach sizeable values on the order of $e/4\pi$ for typical values of proximity-induced SOC at room temperature. This behaviour is robust to imperfections in the SSOF even when the vanishing global average condition, $\langle \Phi(x) \rangle=0$, is not exactly met. We verified this with different types of SOC and SSOF spatial patterns $\Phi(x)$. For example, the spin Hall conductivity presented in Fig. \ref{fig:03}(c) is found to vary by less than 10\% in the presence of a spatially uniform Rashba-like SOC as large as 50\% of the SSOF magnitude itself;  see SM for additional details. Moreover, at variance with 2D conductors subject to the usual uniform Rashba effect \cite{dimitrova_spin-hall_2005,milletari_covariant_2017}, the spatial dependence of the Rashba SSOF protects our quantum geometry-driven spin Hall effect from exact cancellations due to impurity-scattering corrections. In fact, a semiclassical conservation law for expectation values involving the spin current can be derived in the vein of Ref. \cite{milletari_covariant_2017} yielding $\langle H_{\textrm{so}}(x)v_{i}^{z}\rangle=0$, with  $\langle ... \rangle$ denoting a quantum and disorder average. For a uniform Rashba field, this relation (which holds in the presence of arbitrary non-magnetic impurity potentials) implies $\langle v_i^z \rangle = 0$ in steady-state conditions and thus $\mathbf{j}_s^{z}=0$. However, in our system, the condition $\langle v_i^z \rangle = 0$ is circumvented due to the oscillatory nature of $ H_{\textrm{so}}(x)$. The SSOF-driven SHE thus appears to be more robust than its counterpart in standard Rashba-coupled graphene. 
    
We now briefly address the case of 2D metamaterials with concurrent Rashba-type and valley-Zeeman SSOFs. Here, the condition $\langle\Phi(x)\rangle=0$ could be achieved by alternating the relative rotation angle of consecutive TMD layers, exploiting the anti-periodicity of the valley-Zeeman effect, $\lambda_{\textrm{vz}}(\theta)=-\lambda_{\textrm{vz}}(\theta\pm\pi/3)$ \cite{Alessandro_19,Li_19}. The ensuing SSOF, in this case, strongly renormalises the group velocity of wavepackets that propagate parallel to the SSOF direction. The leading correction to Eq. (\ref{eq:En3-perturbation-theory}) is given by $\delta\varepsilon_{(n=\pm 1),{\bf k}s}=\pm\xi\Lambda_{\textrm{vz}}{\,\sin^{2}\theta_{\mathbf{k}}}$,
with $\Lambda_{\textrm{vz}}=\lambda_{\textrm{vz}}^{2}/(\hbar v{G_{1}^{2})}$, yielding an anisotropic dispersion and SB curvature at low energies (see SM). The valley-Zeeman SSOF leads to an overall decrease in the SB curvature magnitude, which is reflected in the spin Hall conductivity [Fig. (\ref{fig:03})(c)]. This is also at odds with the expected behaviour in (standard) proximitised graphene, where the spin Hall conductivity has a non-monotonic behaviour with $\lambda_{\text{vz}}$, with $\lambda_{\text{vz}} \neq 0$ \cite{milletari_covariant_2017,Perkins_SHE_2024} being essential to observe the SHE.

    In closing, we have shown that the spatial patterning of symmetry-breaking spin-orbit fields gives rise to rich physics beyond that of conventional superlattices, in particular, the emergence of 2D massless Dirac fermions with anomalous electrodynamic responses. The proposed periodic modulation of interface-induced SOC is within reach of current nano-fabrication methods and is likely to have broad applications beyond those described in this work. 
  
      The authors acknowledge Samuel Bladwell for  bringing to their attention Refs. \cite{Waveguide_1,Waveguide_2}, and for fruitful discussions on the spatial modulation of SOC in waveguides.  A.F. is grateful for the hospitality of the Faculdade de Engenharia da Universidade do Porto, where most of the work was done. A.F. was partially supported by the Royal Society (U.K.) through a Royal Society University Research Fellowship.
    
%


 
      \clearpage

 \onecolumngrid
\section*{Supplementary Material}

\vspace{1cm}
 
    
\section{Preliminaries}
\subsection{Model and numerical approach}

The low-energy Hamiltonian in the valley-isotropic form is 
\vspace{3mm}
\begin{align}
H_{\tau}=v\,(\boldsymbol{\sigma}\cdot{\bf \mathbf{p}})\otimes s_{0}+U(x)\,\sigma_{0}\otimes s_{0}+H_{\textrm{R}}(x)+H_{\textrm{vz},\tau}(x)+H_{\textrm{KM}}(x), 
\label{eq:ham-all-1}
\\ \notag 
\end{align}
with $U(x)=u\Phi(x)$, $H_{\textrm{R}}(x)=\lambda_R \Phi(x)\,(\sigma_{x}\otimes s_{y}-\sigma_{y}\otimes s_{x})$, $H_{\textrm{vz,\ensuremath{\tau}}}(x)=\tau\lambda_{\textrm{vz}}\Phi(x)\,\sigma_{0}\otimes s_{z}$, and $H_{\textrm{KM}}(x)=\lambda_{\textrm{KM}}\Phi(x)\,\sigma_{z}\otimes s_{z}$  (see main text for definitions). The wavefunctions in valley $\tau=\pm 1$ are 4-component spinors of the form $\Psi_{\tau}({\bf r})=(\psi_{\tau\uparrow}^{A}({\bf r}),\psi_{\tau\downarrow}^{A}({\bf r}),\psi_{\tau\uparrow}^{B}({\bf r}),\psi_{\tau\downarrow}^{B}({\bf r}))^{\textrm{t}}$ for $\tau=1$ and $\Psi_{\tau}({\bf r})=(-\psi_{\tau\uparrow}^{B}({\bf r}),-\psi_{\tau\downarrow}^{B}({\bf r}),\psi_{\tau\uparrow}^{A}({\bf r}),\psi_{\tau\downarrow}^{A}({\bf r}))^{\textrm{t}}$ for $\tau=-1$. Moving to reciprocal space, the eigenproblem formally reduces to solving an infinite set of coupled equations for the plane-wave amplitudes $\{\psi_{\mathbf{k}s}^{\sigma}\}$ for each valley:
\vspace{3mm}
\begin{equation}
\hbar v|\mathbf{k}|e^{-i\sigma\theta_{\mathbf{k}}}\psi_{\mathbf{k}s}^{-\sigma}+\sum_{p\in\mathbb{Z}}\left[\left(u+s\sigma\lambda_{\textrm{KM}}+s \tau \lambda_{\textrm{vz}}\right)\Phi_{G_{p}}\psi_{\mathbf{k}-\mathbf{G}_{p},s}^{\sigma}+i(s-\sigma)\lambda_{\textrm{R}}\Phi_{G_{p}}\psi_{\mathbf{k}-\mathbf{G}_{p},-s}^{-\sigma}\right]=E\psi_{\mathbf{k}s}^{\sigma}\,,
\label{eq:coupled-eqs}
\end{equation}
where $s=\pm\,(\equiv\uparrow,\downarrow)$ and $\sigma=\pm\,(\equiv A,B)$ are the spin and pseudospin indices, respectively; the valley index is omitted for brevity. Furthermore, $\mathbf{k}$ is the Bloch wavevector from the Dirac point, $\theta_{\mathbf{k}}=\angle(\mathbf{k},\hat{x})$, $\mathbf{G}_{p}=G_{p}\hat{x}$ with $G_{p}=2\pi p/a_\text{S}$ ($p\in\mathbb{Z}$), and $\Phi_{G_{p}}$ are the Fourier coefficients of the periodic modulation. 

\vspace{2mm}

The summation over Fourier components in Eq. (\ref{eq:coupled-eqs}) is truncated to a finite, but large, number of terms ($|p|\le N$). The resulting system of coupled equations is solved numerically yielding $d=4(2N+1)$ bands \{${\varepsilon_{n\mathbf{k}}}$\} and associated 4-component eigenvectors \{${\psi_{n\mathbf{k}}}$\}. 

\vspace{2mm}

The spin Berry curvature  of each band $n$ is calculated from  \cite{xiao_berry_2010}
\begin{equation}
\Omega_{n}^{z}(\mathbf{k})=-2\hbar^{2}v^{2}\,\text{Im}\;\sum_{m\neq n}\frac{\psi_{n\mathbf{k}}^{\dagger}\,\hat{\sigma}_{y}\otimes\hat{s}_{z}\,\psi_{m\mathbf{k}}\times\psi_{m\mathbf{k}}^{\dagger}\,\hat{\sigma}_{x}\otimes s_{0}\,\psi_{n\mathbf{k}}}{\left(\varepsilon_{n\mathbf{k}}-\varepsilon_{m\mathbf{k}}\right)^{2}}.
\end{equation}
The linear-response intrinsic spin-Hall conductivity is $\sigma_{yx}^{z}=(e/2) \sum_{\mathbf{k}}\sum_{n}f(\varepsilon_{n\mathbf{k}})\,\Omega_{n}^{z}(\mathbf{k})$, where $f(\varepsilon)$ is the Fermi-Dirac distribution function.   

\subsection{Spatial profile of the periodic perturbation}
We consider Kronig-Penney (KP) and sinusoidal perturbations with zero spatial average. The KP profile is
\begin{align}
\Phi(x) & = 2\Phi\sum_{m=-\infty}^{\infty}\left(R(x+ma_\text{S})-r\right),\ \label{eq:Phi(x)}\,\,\,\,\,\,\,\,\,\,0<r<1,
\end{align}
where  $\Phi$  is the SSOF amplitude and $R(x)=\Theta(x+\ell/2)\Theta(\ell/2-x)$. Here, $\Theta$ is the Heaviside step function, $a_\text{S}$ is the lattice width, $\ell$ is the barrier width ($\ell<a_\text{S}$), and
$r=\ell/a_\text{S}$  (for a square wave $r=0.5$). For pure sinusoidal modulations, we use   $\Phi(x)=\Phi_{c}+\Phi\cos(G_{1}x).$ 

\newpage
 
\section{Electronic structure}
\subsection{Numerics}

\begin{figure}[H]
\begin{centering}
\begin{tabular}{ll}
 \textbf{(a)} &  \textbf{(b)}\tabularnewline
\includegraphics[scale=0.55]{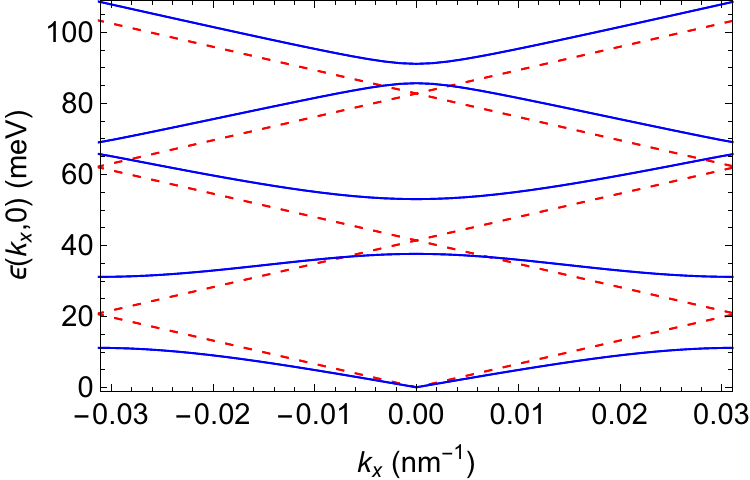} & \includegraphics[scale=0.55]{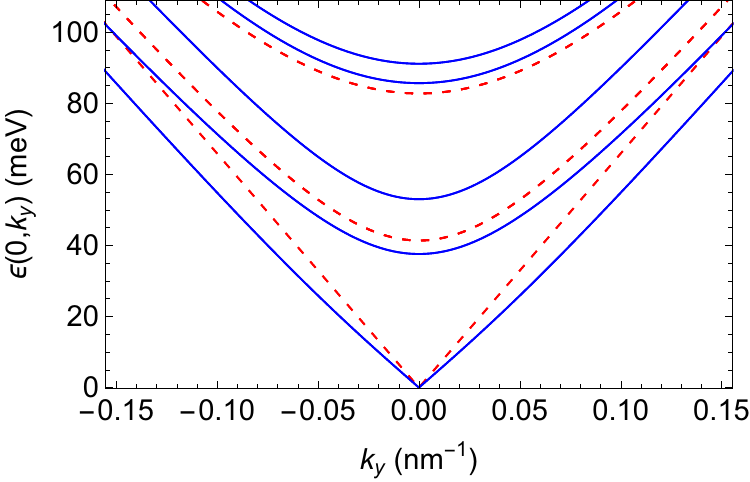}\tabularnewline
\end{tabular}
\par\end{centering}
\begin{centering}
\begin{tabular}{ll}
 \textbf{(c)} &  \textbf{(d)}\tabularnewline
\includegraphics[scale=0.55]{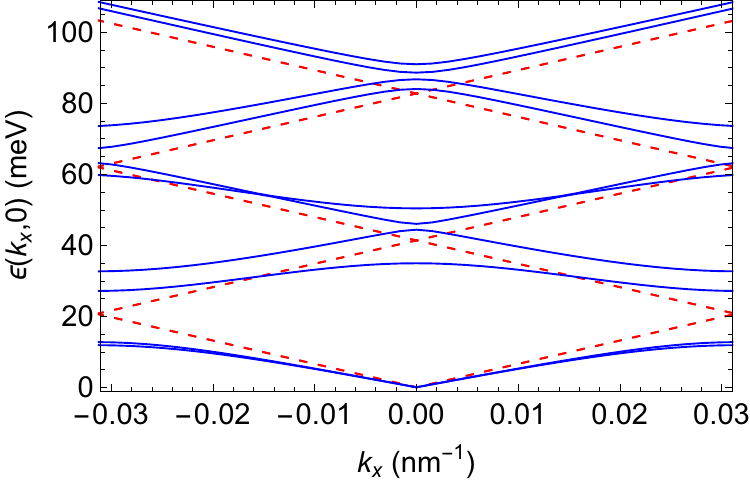} & \includegraphics[scale=0.55]{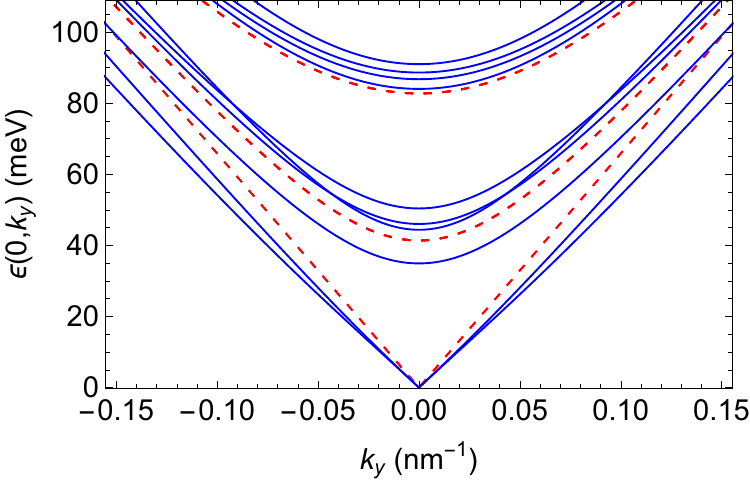}\tabularnewline
\end{tabular}
\par\end{centering}
\caption{Mini-band energy dispersion  $\varepsilon(k_{x},0)$ \textbf{(a,c)} 
and $\varepsilon(0,k_{y})$ \textbf{(b,d)} of graphene subject to a pure Rashba SSOF in the range $k_x \in [-\pi/a_\text{S},\pi/a_\text{S}]$; only positive energy branches are shown. 
SSOF spatial profile: square wave in \textbf{(a,b)} and KP lattice with $r=0.3$ in \textbf{(c,d)}. The spin degeneracy of the bare system without SOC remains intact under square-wave SSOFs [\textbf{(a,b)}], while it is lifted in \textbf{(c,d)}. Other parameters: $a_{\text{S}}=100$ nm and $\lambda_{\text{R}}=20$ meV.
Bare energy dispersion is represented by dashed red lines as a guide to the
eye.}
\label{fig_supp:01}
\end{figure}

\begin{figure}[H]
\begin{centering}
\begin{tabular}{ll}
 \textbf{(a)} &  \textbf{(b)}\tabularnewline
\includegraphics[scale=0.45]{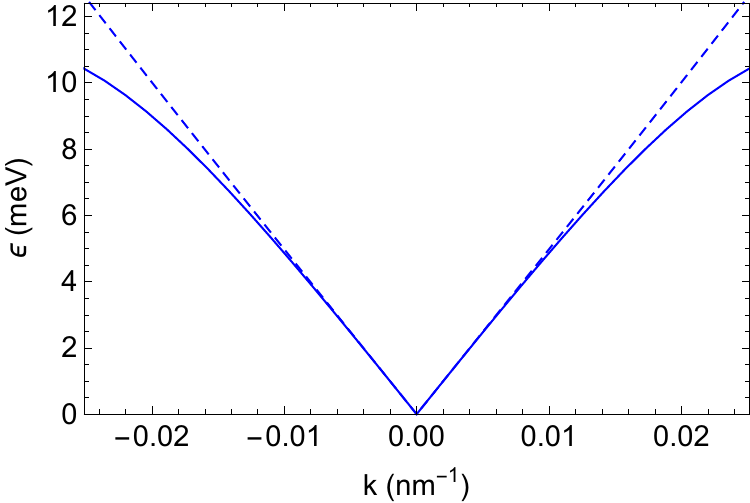} & \includegraphics[scale=0.45]{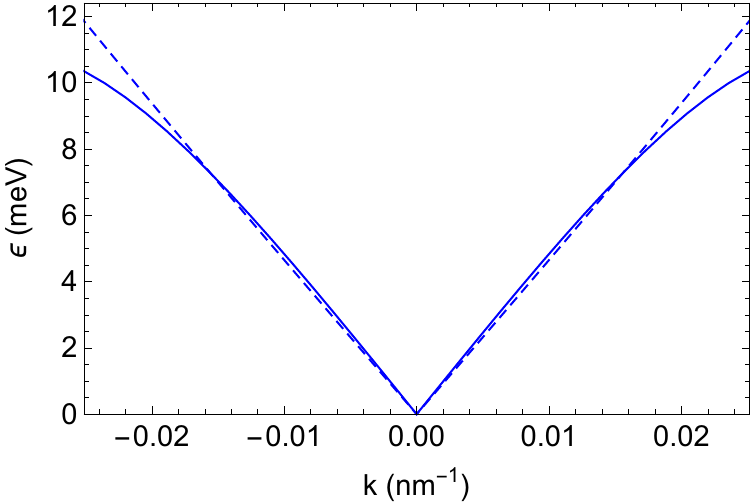}\tabularnewline
\end{tabular}
\par\end{centering}
\caption{Low-energy dispersion for a pure Rashba SSOF \textbf{(a)}
and a mixed Rashba-valley-Zeeman SSOF \textbf{(b)}. Solid and dashed lines show $\varepsilon(k_{x},0)$ and $\varepsilon(0,k_{y})$, respectively. The low-energy Dirac cones are isotropic in (a) and anisotropic in (b).
Parameters: $r=0.5$, $a_{\text{S}}=100$ nm, $\lambda_{\text{R}}=20$ meV (a,b)   and $\lambda_{\text{vz}}=10$ meV (b).}
\label{fig_supp:03}
\end{figure}

Figure \ref{fig_supp:01} shows the numerically calculated electronic structure of graphene subject to a \textit{pure Rashba SSOF} for square-wave and KP modulations. We note that only positive-energy states are shown for clarity. Main findings are:

\begin{itemize}

  \item spin-degenerate Dirac cones emerge at low energy (for $r=0.5$);
  \item the Dirac cones are isotropic around $\bf k = 0$ (see also Fig. \ref{fig_supp:03}(a)); 
  \item the mini-bands at higher energy are separated by pseudo-gaps;
  \item spin splittings are visible for KP profiles with $r\neq 0$ (see panels (c) and (d)).
    
\end{itemize}

\vspace{0.5cm}

\vspace{1cm}

We also verified that sine-wave SSOFs behave similarly to square-wave SSOFs but with less pronounced features. This is easily explained by comparing their Fourier coefficients. For a square wave profile, $\Phi_{\pm G_{1}}^{\textrm{square}}=(2/\pi)\Phi$, while for sine waves: $\Phi_{\pm G_{1}}^{\textrm{sine}}=(1/2)\Phi$. 

\newpage

\begin{figure}[H]
\begin{centering}
\begin{tabular}{ll}
 \textbf{(a)} &  \textbf{(b)}\tabularnewline
\includegraphics[scale=0.55]{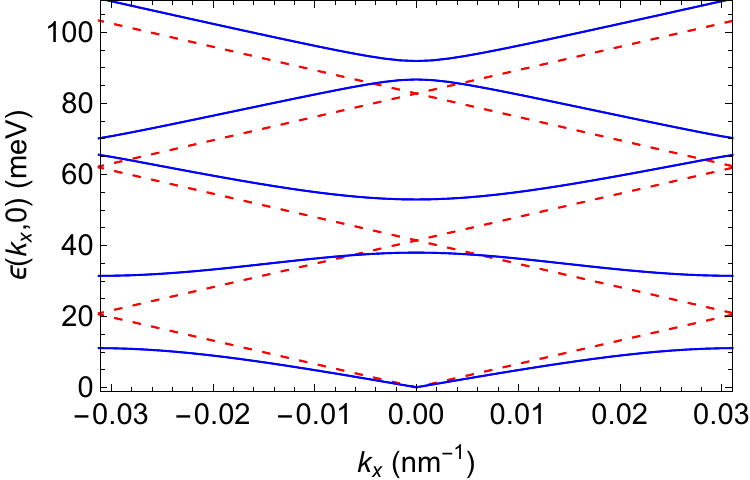} & \includegraphics[scale=0.55]{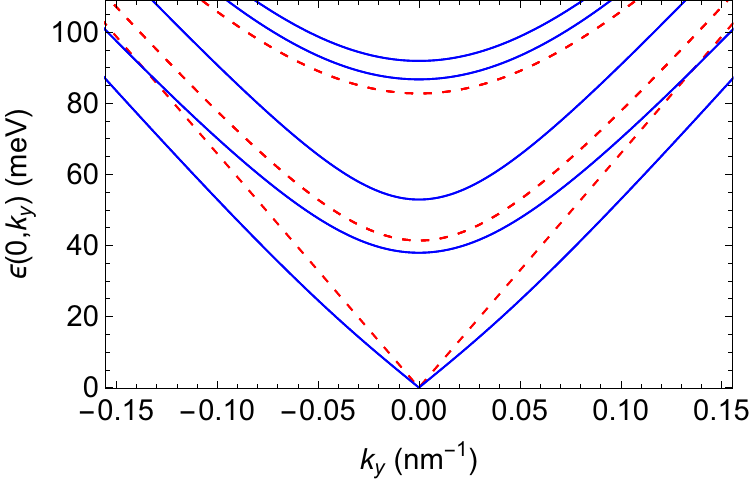}\tabularnewline
\end{tabular}
\par\end{centering}
\begin{centering}
\begin{tabular}{ll}
 \textbf{(c)} &  \textbf{(d)}\tabularnewline
\includegraphics[scale=0.55]{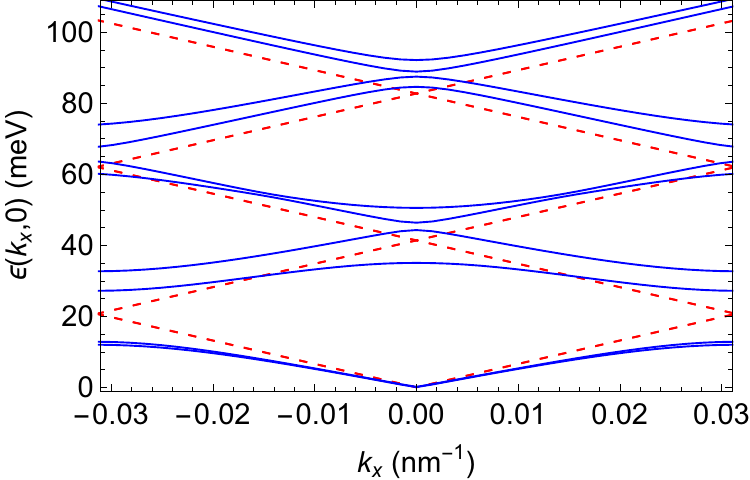} & \includegraphics[scale=0.55]{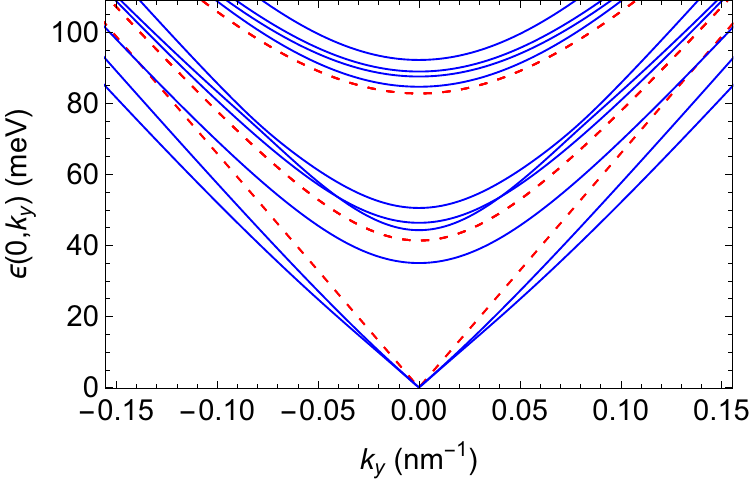}\tabularnewline
\end{tabular}
\par\end{centering}
\caption{Mini-band energy dispersion  $\varepsilon(k_{x},0)$ \textbf{(a,c)} 
and $\varepsilon(0,k_{y})$ \textbf{(b,d)} for a mixed Rashba-valley-Zeeman SSOF.
SSOFs spatial profile: square wave in \textbf{(a,b)} and KP lattice with $r=0.3$ in \textbf{(c,d)}. The valley-Zeeman strength is $\lambda_{\text{vz}}=10$ meV. Other parameters as in Fig. \ref{fig:01}.}
\label{fig_supp:02}
\end{figure}

The electronic structure for mixed \textit{Rashba-valley-Zeeman SSOFs} is shown in Fig. \ref{fig_supp:02}. The main difference with respect to Fig. \ref{fig_supp:01} is that the emergent Dirac cones become anisotropic irrespective of the SSOF's spatial profile. 

\subsection{Perturbation theory}
We start with the class of $2\times 2$ Dirac Hamiltonians subject to a periodic perturbation (period $a_\text{S}$) with generic pseudospin structure  of the type:
\medskip{}
\begin{equation}
\hat{H}=\hbar v(\sigma_{x}\hat{k}_{x}+\sigma_{y}\hat{k}_{y})+\hat{S}(x)+\hat{X}(x)+\hat{Y}(x)+\hat{Z}(x),
\\ \notag 
\label{eq:hamiltonianSXYZ}
\end{equation}
where $\hat{S}(x)=S\Phi(x)\sigma_{0}$ is the superlattice potential and $\hat{X}(x)=X\Phi(x)\sigma_{x}$,
$\hat{Y}(x)=Y\Phi(x)\sigma_{y}$ and $\hat{Z}(x)=Z\Phi(x)\sigma_{z}$ are the  SU(2)-pseudospin fields.
We recall that the unperturbed eigenstates are the spinors $\psi_{{\bf k}}^{0\sigma}({\bf r})=\frac{1}{\sqrt{2}}(1,\sigma e^{i\theta_{{\bf k}}})^{\text{t}}e^{i{\bf k}\cdot{\bf r}}$ corresponding to the
eigenenergies $\varepsilon_{{\bf k}\sigma}^{0}=\sigma\hbar v |{\bf k}|$. Using second order perturbation theory, we find that that the correction to the lowest-lying energy states, $\delta \varepsilon_{\mathbf{k}\sigma}^{(2)}=\varepsilon_{\mathbf{k}\sigma}^{(2)} -\varepsilon_{{\bf k}\sigma}^{0} $  is
\medskip{}
\begin{align*}
\delta \varepsilon_{\mathbf{k}\sigma}^{(2)}= \sigma\frac{1}{\hbar v}\sum_{n\neq0}\Bigg[\frac{2|{\bf k}|-|{\bf G}_{n}|\cos\theta_{{\bf k},{\bf G}_{n}}}{|{\bf k}|^{2}-|{\bf k}-{\bf G}_{n}|^{2}}S_{G_{n}}^{2}+\frac{|{\bf G}_{n}|\cos\theta_{{\bf k},{\bf G}_{n}}}{|{\bf k}|^{2}-|{\bf k}-{\bf G}_{n}|^{2}}Z_{G_{n}}^{2}
\end{align*}
\begin{align*}
 & +\frac{2|{\bf k}|\cos^{2}\theta_{{\bf k}}-|{\bf G}_{n}|\cos\theta_{\mathbf{k},{\bf {\bf G}}_{n}}\cos\theta_{{\bf k}}}{|{\bf k}|^{2}-|{\bf k}-{\bf G}_{n}|^{2}}X_{G_{n}}^{2}+\frac{2|{\bf k}|\sin^{2}\theta_{{\bf k}}+|{\bf G}_{n}|\cos\theta_{\mathbf{k},{\bf {\bf G}}_{n}}\cos\theta_{{\bf k}}}{|{\bf k}|^{2}-|{\bf k}-{\bf G}_{n}|^{2}}Y_{G_{n}}^{2}
\end{align*}
\begin{align*}
 & +\frac{2|{\bf k}|\cos\theta_{{\bf k}}-|{\bf G}_{n}|\cos\theta_{\mathbf{k},{\bf G}_{n}}}{|{\bf k}|^{2}-|{\bf k}-{\bf G}_{n}|^{2}}S_{G_{n}}X_{G_{n}}+\frac{2|{\bf k}|\sin\theta_{{\bf k}}}{|{\bf k}|^{2}-|{\bf k}-{\bf G}_{n}|^{2}}S_{G_{n}}Y_{G_{n}}
\end{align*}
\begin{align*}
 & +\frac{4|{\bf k}|\cos\theta_{{\bf k}}\sin\theta_{{\bf k}}+2|{\bf G}_{n}|\cos\theta_{\mathbf{k},{\bf G}_{n}}\sin\theta_{{\bf k}}}{|{\bf k}|^{2}-|{\bf k}-{\bf G}_{n}|^{2}}X_{G_{n}}Y_{G_{n}}\Bigg].
 \label{eq:spectrum-SXYZ-2nd-order}
\end{align*}
Expanding around the $K$ point (${\bf k}={\bf 0}$) up to first order in $\bf k$, we obtain
\medskip{}
\begin{eqnarray}
\varepsilon_{{\bf k}\sigma}^{(2)} & \approx & \sigma\hbar v\left[1-\xi\frac{\sin^{2}\theta_{{\bf k}}S^{2}+\cos^{2}\theta_{{\bf k}}Z^{2}+Y^{2}+\sin\theta_{{\bf k}}SY+2\sin\theta_{{\bf k}}\cos\theta_{{\bf k}}XY}{\hbar{{}^2}v{{}^2}G_{1}{{}^2}}\right]|{\bf k}|.
\label{eq:energy-0xyz-2nd-order}
\\ \notag 
\end{eqnarray}
where $\xi$ is a geometric factor equal to $1$ for
a sine-wave modulation and $\sum_{n=1}^{\infty}(4r/n){{}^2}{\rm sinc{{}^2}}(n\pi r)$
for a KP profiles with zero spatial average. We note that the gapless nature of the spectrum
has its origin in the chiral phase factors $e^{i\theta_{{\bf k}}}$
of the unperturbed eigenstates as first discussed in Ref. \cite{SL_Park_08} for pure scalar perturbations. 

\medskip{}
\medskip{}
Generalizing these results to the $4\times 4$ Dirac Hamiltonian, containing the SSOFs, yields
\medskip 

\begin{equation}
\varepsilon_{{\bf k}s\sigma}^{(2)}\approx
\sigma\hbar v\Bigg[1-
\mu\frac{\sin^{2}\theta_{{\bf k}}(u^{2}+\lambda_{\textrm{vz}}^2)
+\cos^{2}\theta_{{\bf k}}\lambda_{\textrm{KM}}^{2}
+\lambda_{\text{R}}^{2}
+2s \sqrt{\sin^{2}\theta_{{\bf k}}u^{2} (\lambda_{\textrm{vz}}^2+\lambda_{\text{R}}^2) 
+\cos^{2}\theta_{{\bf k}} 
\lambda_{\textrm{KM}}^2\lambda_{\text{R}}^2 }}{\hbar^{2}v^{2}G_{1}^{2}}\Bigg]\ |{\bf k}|.
\label{eq:2nd.order-VUW}
\end{equation}

Equation (\ref{eq:2nd.order-VUW}) provides the following insights:
\begin{itemize}

\item \textbf{Symmetry protection}: The Dirac point degeneracy is robust against all periodic perturbations with $\langle \Phi(x)\rangle =0$, irrespective of the local spatial symmetries broken by the SSOF. A gap at $\textbf{k}=0$ can only be open by adding uniform SOC terms (i.e. for spatial profiles $\Phi(x)$ witn non-zero spatial average). 
  \item \textbf{Spin splitting}: Surprisingly, a particular type of SSOF \textit{cannot} perturbatively lift the spin degeneracy of the low-lying states on its own. Only specific admixtures of periodic perturbations can lead to spin splitting of low-lying states (e.g., Rashba SSOF combined with a periodic potential).  
  \item \textbf{Emergent Dirac cones}: Pure Rashba SSOFs generate \textit{isotropic Dirac cones} around $\bf k = 0$. Because the spin degeneracy is intact in this case, the low-lying spectrum around the Dirac points mimic that of pristine graphene with a renormalized Fermi velocity. 
  \item{\textbf{Anisotropic behavior}: For SSOFs admixing Rashba and valley-Zeeman components, for example, or in the presence of a periodic potential, $u\neq0$, the energy dispersion becomes anisotropic.} 
  
\end{itemize}

The  group velocity for graphene under the combined action of a Kane-Mele SSOF and a periodic potential $U(x)$ of the main text   can be directly derived from Eq. (\ref{eq:2nd.order-VUW}) by setting $\lambda_{\textrm{R}}=\lambda_{\textrm{vz}}=0$.

\subsection{Third-order perturbation theory versus numerics}
Our fully non-perturbative numerical calculations demonstrate that pure Rashba SSOFs lift the spin degeneracy of mini-bands for SSOFs with $r \neq 0.5$; see Fig. \ref{fig_supp:02}. This effect is significant away from the Dirac point.  To capture this feature, one must turn to third-order perturbation theory. After tedious algebra, the energy dispersion to leading order in $\bf k$ and associated spin gap, $\Delta_{\text{spin}}(\mathbf{k})$, are obtained as 
\begin{align}
\varepsilon_{{\bf k}s\sigma}^{(3)}\approx\varepsilon_{{\bf k}\sigma}^{(0)}-\xi\sigma\frac{\lambda_{\text{R}}^{2}}{\hbar vG_{1}^{2}}\left(|{\bf k}|\,-s\frac{\tilde{\xi}}{\xi}\frac{\lambda_{\text{R}}}{\hbar vG_{1}^{2}}\ |{\bf k}|^{2}\right),\,\,\,\,\,\,\,\,\,\,\, \Delta_{\text{spin}}(\mathbf{k})=2\tilde{\xi}|\lambda_{\textrm{R}}|\left(\frac{\lambda_{\text{R}}}{\hbar vG_{1}^{2}}\right)^{2}\ |{\bf k}|^{2},
\label{eq:3rd.order-W}
\end{align}
where $\tilde{\xi}$ is a geometric factor, $\tilde{\xi}=\sideset{}{_{n,m}^{\prime}}\sum4(2r)^{3}{\rm sinc}(n\pi r){\rm sinc}(m\pi r){\rm sinc}((n+m)\pi r)/[n^{2}(m+n)^{2}]$ (the primed summation is used to exclude the cases $n=-m$ and $n,m=0$). In accord with the numerics, we see that $\tilde{\xi}=0$ for
$r=0.5$ (spin-degenerate states) and $\tilde{\xi}\neq0$ for $r\neq0.5$. The perturbative results
reproduce well the numerics within their regime of validity, $|u|,|\lambda_\text{KM}|,|\lambda_{\text{R}}|,|\lambda_{\text{vz}}|\ll\pi\hbar v/a_\text{S}$
($\approx20$ meV for $a_\text{S}=100$ nm). 
\begin{figure}[H]
\begin{centering}
\begin{tabular}{ll}
(a) & (b)\tabularnewline
\includegraphics[scale=0.55]{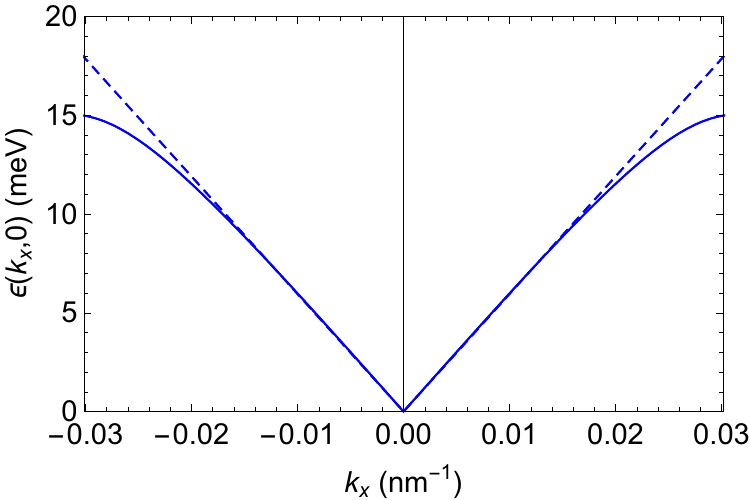} & \includegraphics[scale=0.55]{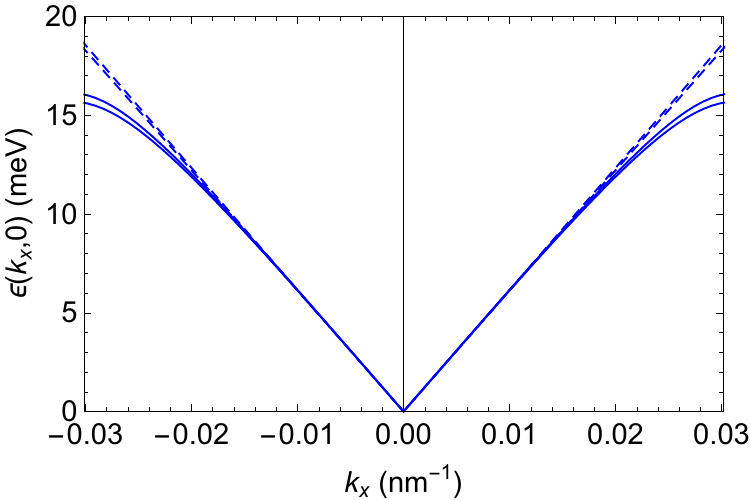}\tabularnewline
\end{tabular}
\par\end{centering}
\caption{\textit{Perturbation theory versus numerics}: Energy dispersion  $\varepsilon(k_{x},0)$ of the lowest-lying electron band   for a pure Rashba SSOF with $r=0.5$ \textbf{(a)} [$r=0.3$\textbf{ (b)}], obtained numerically (solid lines) and via Eq. (\ref{eq:3rd.order-W}) (dashed lines). Other parameters:  $a_{\text{S}}=100$ nm and $\lambda_\text{R}=10$ meV.}
\label{fig_supp:04}
\end{figure}

\subsection{Zero-energy solutions: analytical results}
 
 Here, we provide explicit solutions for the $K$ valley (note that the expressions for $\tau=-1$ can be obtained from those below by simply inverting the sign of $\lambda_{\text{sv}}$). The spinor components can be cast as $\psi_{s}^{\sigma}({\bf r})=\phi_{s}^{\sigma}(x)e^{ik_{y}y}$, where $k_y$ is the wavevector component perpendicular to the superlattice direction. The Dirac equation for zero-energy states ($\hbar \equiv 1$) yields
\medskip{}
\begin{equation}
-i\hbar v\partial_{x}\phi_{s}^{-\sigma}(x)+\left(u+s\sigma \lambda_\text{KM}+s \lambda_\text{vz}\right)\Phi(x)\phi_{s}^{\sigma}(x)+i(s-\sigma)\lambda_\text{R}(x)\phi_{-s}^{-\sigma}(x)=0\,.
\label{eq:zero-energy-simplified}
\end{equation}
Equation (\ref{eq:zero-energy-simplified}) admits analytical solutions in a number of cases summarized below. We write these solutions in terms of the function 
$\varphi_{\beta}(x)=\beta/v \int^{x}\Phi(x^{\prime})dx^{\prime}$. Moreover, $L$ is the linear dimension 
of the system along $x$ $(L\gg a_\text{S})$. 
\medskip{}
\begin{itemize}

  \item \textbf{Case 1:\textit{ No Rashba field}} ($\lambda_\text{R}=0$, $\lambda_\text{KM}\neq 0$, 
  $\lambda_{\text{vz}}\neq 0$, $u \neq 0$). In this case, up and down spin sectors decouple. The zero-energy eigenfunctions for $s=\uparrow,\downarrow$ may be written as

  \begin{equation}
   \phi^\sigma_{s}(x)=\frac{1}{\sqrt{L}}  \left[\begin{array}{c}
    A\cos\varphi_{\gamma}(x)+iB\sin\varphi_{\gamma}(x)\\
    -B\gamma/\Delta\gamma\cos\varphi_{\gamma}(x)-iA\gamma/\Delta\gamma\sin\varphi_{\gamma}(x)
    \end{array}\right],
  \label{eq:rashbazero}
  \end{equation}
  
where $\gamma=((u+\lambda_\text{vz})^{2}-\lambda_\text{KM}^{2})^{1/2}$, 
$\Delta\gamma=u+\lambda_\text{vz}-\lambda_\text{KM}$,
and $A$ and $B$ are constants. There are two degenerate solutions for a each spin state. We present two examples.
For $u \neq 0$ and all other potentials equal to zero, two orthogonal degenerate zero-energy eigenstates for ${\bf k}=0$ corresponding to the two zero-energy modes (ZEMs)  read 

\begin{equation}
   \phi^A_{s}(x)=\frac{1}{\sqrt{L}} \left[\begin{array}{c}
    \cos\varphi_{u}(x) \\
    -i\sin\varphi_{u}(x)
\end{array}\right],
  \end{equation}

\begin{equation}
   \phi^B_{s}(x)=\frac{1}{\sqrt{L}} \left[\begin{array}{c}
    -i\sin\varphi_{u}(x) \\
    \cos\varphi_{u}(x)
\end{array}\right],
  \end{equation}

while for $\lambda_\text{KM}\neq 0$ and all the other potentials zero, the ZEMs are 

\begin{equation}
   \phi^A_{s}(x)=\frac{1}{\sqrt{C_{\lambda_\text{KM}}}} \left[\begin{array}{c}
    \cosh\varphi_{\lambda_\text{KM}}(x) \\
    -i\sinh\varphi_{\lambda_\text{KM}}(x)
\end{array}\right],
  \end{equation}

\begin{equation}
   \phi^B_{s}(x)=\frac{1}{\sqrt{C_{\lambda_\text{KM}}}} \left[\begin{array}{c}
    i\sinh\varphi_{\lambda_\text{KM}}(x) \\
    \cosh\varphi_{\lambda_\text{KM}}(x)
\end{array}\right],
  \end{equation}
where $C_{\lambda_\text{KM}}=\int_{-L/2}^{L/2}dx\cosh 2\varphi_{\lambda_\text{KM}}(x)$. 
 Eq. (\ref{eq:rashbazero}) holds for
$u+\lambda_\text{vz}\neq\lambda_\text{KM}$. When
$u+\lambda_\text{vz}=\lambda_\text{KM}=\eta$, the
 ZEMs are

\begin{equation}
   \phi^A_{s}(x)=\frac{1}{\sqrt{C_\eta}} 
    \left[\begin{array}{c}
    1 \\
    -2i\varphi_{\eta}(x)
          \end{array}\right],
  \end{equation}

\begin{equation}
   \phi^B_{s}(x)=\frac{1}{\sqrt{L}} \left[\begin{array}{c}
    0 \\
    1
\end{array}\right],
  \end{equation}
where $C_{\eta}=\int_{-L/2}^{L/2}dx(1+4\varphi_{\eta}^{2}(x))$.

  \item \textbf{Case 2: \textit{Pure Rashba field}} ($\lambda_\text{R}\neq 0$, $\lambda_{\text{KM}}= 0$, 
  $\lambda_{\text{vz}}= 0$, $u = 0$). In this case, 
  four orthogonal degenerate zero-energy eigenstates for ${\bf k}=0$
corresponding to the four ZEMs are found:

  \begin{equation}
   \phi_\uparrow^A(x)=\frac{1}{\sqrt{L}}  \left[\begin{array}{c}
    1 \\
    0 \\
    0 \\
    0
  \end{array}\right],
  \end{equation}

\begin{equation}
   \phi_\uparrow^B(x)=\frac{1}{\sqrt{C_{\lambda_\text{R}}} }
   \left[\begin{array}{c}
    0 \\
    1 \\
    0 \\
    -2\varphi_{\lambda_\text{R}}(x)
  \end{array}\right],
  \end{equation}

\begin{equation}
   \phi_\downarrow^A(x)=\frac{1}{\sqrt{C_{\lambda_\text{R}}} }
   \left[\begin{array}{c}
    2\varphi_{\lambda_\text{R}}(x) \\
    0 \\
    1 \\
    0
  \end{array}\right],
  \end{equation}

\begin{equation}
   \phi_\downarrow^B(x)=\frac{1}{\sqrt{L}}
   \left[\begin{array}{c}
    0\\
    0 \\
    0 \\
    1
  \end{array}\right],
  \end{equation}

where $C_{\lambda_\text{R}}=\int_{-L/2}^{L/2}dx(1+4\varphi_{\lambda_\text{R}}^{2}(x))$.

\end{itemize}

These analytical results highlight the robustness of the Dirac point degeneracy to SSOFs satisfying the zero spatial-average condition, in accord with the exact numerical calculations.  

\section{Impact of spatially uniform components of SOC}

\begin{figure}[H]
\begin{centering}
\begin{tabular}{ll}
\textbf{(a)} &  \textbf{(b)}\tabularnewline
\includegraphics[scale=0.55]{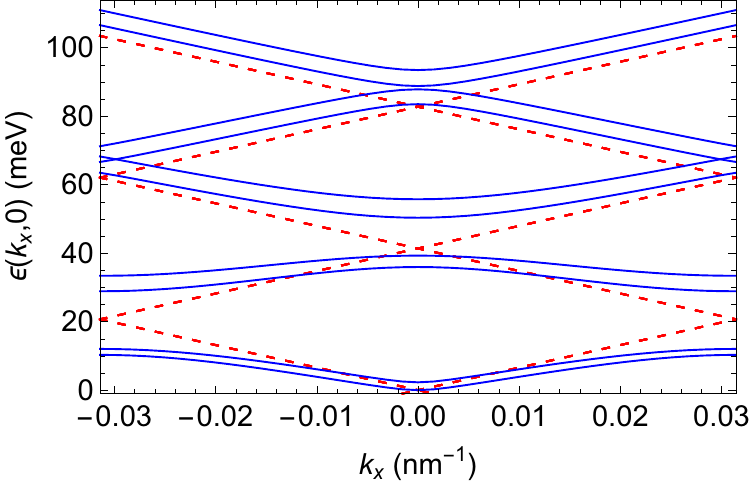} & 
\includegraphics[scale=0.55]{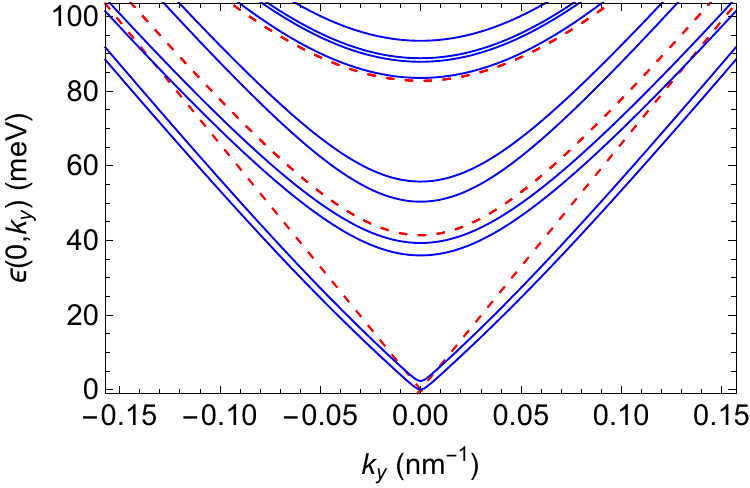}\tabularnewline
\end{tabular}
\par\end{centering}
\caption{Mini-band energy dispersion $\varepsilon(k_{x},0)$ \textbf{(a)} 
and $\varepsilon(0,k_{y})$ \textbf{(b)} of graphene subject to a square wave ($r=0.5$)
Rashba modulation ($\lambda_{\text{R}}=20$ meV) with a uniform component corresponding to 
$\lambda_{\textrm{Rc}}=2$ meV; only positive energy branches are shown.
Other parameters: $a_{\text{S}}=100$ nm. Bare energy dispersion is represented by dashed red lines as a guide to the
eye.}
\label{fig_supp:05}
\end{figure}

Figure \ref{fig_supp:05} shows the numerically calculated electronic structure of a graphene subject to a Rashba SSOF with non-zero global average, 
$\langle \Phi(x)\rangle \neq 0$. Compared to the case of a Rashba SSOF with $\langle \Phi(x)\rangle = 0$ discussed in detail in the main text [see also panels 
\textbf{(a)} and \textbf{(b)} of Fig. \ref{fig_supp:01} above, we see that the main modifications introduced by the spatially uniform Rashba component, $\lambda_{\textrm{Rc}}$, are twofold:

\begin{itemize}

  \item lifting of spin degeneracy
  \item suppression of Dirac cones in favour of massive low-energy mini-bands.
  
\end{itemize} 
\vspace{0.5cm}

\begin{figure}[H]
\begin{centering}
\begin{tabular}{ll}
\textbf{(a)} &  \textbf{(b)}\tabularnewline
\includegraphics[scale=0.6]{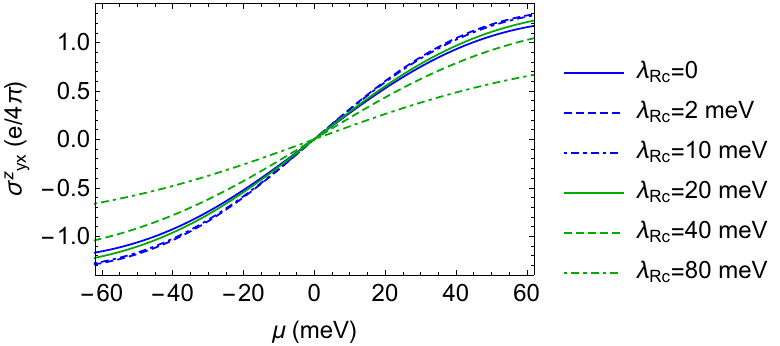} & 
\includegraphics[scale=0.6]{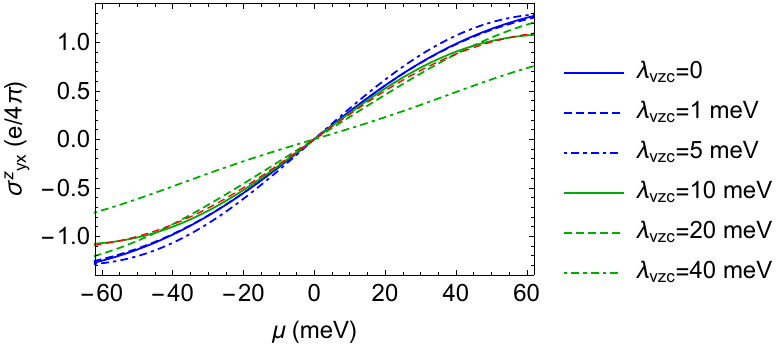}\tabularnewline
\end{tabular}
\par\end{centering}
\caption{Spin-Hall conductivity for square-wave ($r=0.5$) SSOFs with non-zero global average as a function of chemical potential for $k_B T = 26$ meV 
and selected values of the uniform SOC components, $\lambda_{\textrm{Rc}}$ and $\lambda_{\textrm{zvc}}$, as indicated, for two scenarios: \textbf{(a)}: Pure Rashba-type SSOF with $\lambda_{\text{R}}=20$ meV; and \textbf{(b)}: Admixed Rashba--valley-Zeeman SSOF with $\lambda_{\text{R}}=20$ meV,
$\lambda_{\textrm{Rc}}=10$ meV, and $\lambda_{\text{vz}}=10$ meV. Blue solid line of panel \textbf{(a)} and red dashed line of panel \textbf{(b)} correspond to blue solid line and blue dashed line of Fig. 3\textbf{(c)} in the main text, respectively. Other parameters as in Fig. \ref{fig_supp:05}.}
\label{fig_supp:06}
\end{figure}

The impact of a uniform Rashba component on the room-temperature spin-Hall conductivity is shown in Fig. \ref{fig_supp:06} (a). We see that the spin-Hall response has 
a non-monotonic behavior with $\lambda_{\text{Rc}}$, with $|\sigma_{yx}^z|$ increasing slowly with $\lambda_{\text{Rc}}$
up to $\lambda_{\text{Rc}} \approx \lambda_{\text{R}}/2$ 
and then quickly decreasing beyond that. The response for a system subject to an
admixed Rashba-valley-Zeeman SSOF containing a uniform valley-Zeeman component, $\lambda_{\textrm{vzc}}$, 
is shown in Fig. \ref{fig_supp:06} (b). In this case, we also find that the spin-Hall response shows 
a non-monotonic behavior with $\lambda_{\text{vzc}}$,
with $|\sigma_{yx}^z|$ changing slightly with $\lambda_{\text{vzc}}$
up to $\lambda_{\text{vzc}} \approx 2 \lambda_\text{\text{vz}}$,
while for values much greater it decreases significantly. These results show that $\sigma^z_{yx}$ is robust against  fabrication imperfections leading to small uniform SOC components in the SSOF field.

\end{document}